\documentclass[usenatbib]{mn2e}

\usepackage{amsmath}
\usepackage{graphicx}
\usepackage{natbib}
\usepackage{color}
\usepackage{times}

\newcommand{\snia}{SN~Ia}
\newcommand{\sneia}{SNe~Ia}

\newcommand{\nifs}{$^{56}$Ni}
\newcommand{\chisq}{$\chi^2$}
\newcommand{\dmft}{$\Delta m_{15}(B)$}
\newcommand{\nodata}{. . .}

\newcommand{\siline}{\SiII~$\lambda$6355}
\newcommand{\siblue}{\SiII~$\lambda$3858}
\newcommand{\canir}{\CaII\ NIR triplet}
\newcommand{\cahk}{\CaII\ H\&K}

\newcommand{\vsi}{$v_{\rm Si}$}
\newcommand{\vbsi}{$\bar{v}_{\rm Si}$}
\newcommand{\vbci}{$\bar{v}_{\rm CI}$}
\newcommand{\vbch}{$\bar{v}_{\rm CH}$}
\newcommand{\rhvf}{$R_{\rm HVF}$}
\newcommand{\kms}{km\,s$^{-1}$}

\newcommand{\SiII}{Si~{\sc ii}}

\newcommand{\SII}{S~{\sc ii}}

\newcommand{\CaII}{Ca~{\sc ii}}

\title{High-Velocity Features in Type Ia Supernova Spectra}

\author[Childress, Filippenko, Ganeshalingam, \& Schmidt]{
Michael~J.~Childress$^{1,2}$\thanks{E-mail:mjc@mso.anu.edu.au},
Alexei~V.~Filippenko$^{3}$,
Mohan~Ganeshalingam$^{3,4}$,
\newauthor Brian~P.~Schmidt$^{1,2}$\\
$^{1}$ Research School of Astronomy and Astrophysics, 
Australian National University, 
Canberra, ACT 2611, Australia.\\
$^{2}$ARC Centre of Excellence for All-sky Astrophysics (CAASTRO).\\
$^{3}$Department of Astronomy, University of California, Berkeley, CA 94720-3411, USA.\\
$^{4}$Lawrence Berkeley National Laboratory, Berkeley, CA 94720, USA.
}

\begin{document}
\maketitle

\begin{abstract}
We use a sample of 58 low-redshift ($z \leq 0.03$) Type Ia supernovae (\sneia) having well-sampled light curves and spectra near maximum light to examine the behaviour of high-velocity features (HVFs) in \snia\ spectra. We take advantage of the fact that \siline\ is free of HVFs at maximum light in all \sneia, while HVFs are still strong in the Ca~{\sc ii} near-infrared feature in many SNe, allowing us to quantify the strength of HVFs by comparing the structure of these two lines. We find that the average HVF strength increases with decreasing light-curve decline rate, and rapidly declining \sneia\ ($\Delta m_{15}(B) \geq 1.4$ mag) show no HVFs in their maximum-light spectra. Comparison of HVF strength to the light-curve colour of the \sneia\ in our sample shows no evidence of correlation. We find a correlation of HVF strength with the velocity of \siline\ at maximum light (\vsi), such that \sneia\ with lower \vsi\ have stronger HVFs, while those \sneia\ firmly in the ``high-velocity'' (i.e., \vsi $\geq$ 12,000~\kms) subclass exhibit no HVFs in their maximum-light spectra. While \vsi\ and \dmft\ show no correlation in the full sample of \sneia, we find a significant correlation between these quantities in the subset of \sneia\ having weak HVFs. In general, we find that slowly declining (low \dmft) \sneia, which are more luminous and more energetic than average \sneia, tend to produce {\em either} high photospheric ejecta velocities (i.e., high \vsi) {\em or} strong HVFs at maximum light, but not both. Finally, we examine the evolution of HVF strength for a sample of \sneia\ having extensive pre-maximum spectroscopic coverage and find significant diversity of the pre-maximum HVF behaviour.
\end{abstract}


\section{Introduction}
\label{sec:intro}
Type Ia supernovae (\sneia) remain the best standardisable candles for mapping the expansion history of the Universe and thereby constraining the nature of dark energy \citep{riess98, perlmutter99}.  \snia\ luminosities appear to be governed to first order by an underlying physical mechanism which drives both the peak luminosity of the SN and the decline rate (\dmft) of its light curve \citep{phillips93}.  The observed range of \dmft\ values is likely driven by the amount of \nifs\ produced in the explosion and the opacity of the ejecta \citep{hk96, pe00, mazzali01, mazzali07}, but the driving physical mechanism behind the realised range of these parameters remains unclear. Photometric and spectroscopic behaviour which correlates with \dmft\ can provide important clues to, and constraints on, the physical mechanism driving the decline rate.  

Much recent effort has been devoted to investigating the spectroscopic behaviour of \sneia, and whether features in the \snia\ spectra can elucidate the origin of \snia\ luminosity diversity. Central to these efforts has been the study of the \siline\ line, the most prominently identifiable feature of \snia\ spectra. The ratio of absorption strength of the \siline\ line to that of its neighbouring line \SiII\ $\lambda$5972 has long been known to be sensitive to the SN ejecta temperature and correlates with \dmft\ \citep{nugent95, hachinger08}. The two-dimensional parameter space defined by these lines has been proposed as a diagnostic tool for inspecting \snia\ subclasses \citep{benetti05, branch09}.  More recently, the velocity of the absorption minimum of the \siline\ line (\vsi) in the maximum-light spectra of \sneia\ has been used to divide \sneia\ into two groups, ``normal'' and ``high-velocity'' (HV) \sneia, which exhibit different colour behaviour \citep{wang09} and appear to have different spatial distributions within their host galaxies \citep{wang13}.

Another feature in \snia\ spectra of recent interest is the frequent presence of high velocity features (HVFs) whose velocities are significantly higher (typically by a few thousand \kms) than the remainder of the normal ``photospheric'' absorption features.  These features often appear as double-peaked absorption profiles in \siline\ or the Ca~{\sc ii} near-infrared (NIR) triplet \citep[e.g.,][]{gerardy04, mazzali05a, mazzali05b, wang05cf, foley09ig, childress12fr}, and are consistently stronger in early-time \snia\ spectra.  
The physical origin of the HVFs remains uncertain, but broadly speaking they must be caused by some absorbing material outside the nominal photosphere of the SN. Some suggestions for the origin of HVFs include a general density enhancement caused by swept-up \citep{gerardy04} or distant \citep{tanaka06} circumstellar material, an enhancement in the abundance of intermediate-mass elements (IMEs) in the outermost layers of \snia\ ejecta \citep[see discussion in, e.g.,][]{mazzali05a, mazzali05b, tanaka08}, or variations in the ionization state of IMEs in the outer layers of \snia\ ejecta \citep{blondin13}.

In this work we focus on the relationship between the strength of HVFs in \snia\ spectra at maximum light and other properties of the SN, most importantly \dmft\ and \vsi.  We exploit the fact that HVFs in the \siline\ line have universally faded in all \sneia\ by maximum light but HVFs in the \canir\ are still strong in many SNe. The \siline\ line gives an independent measurement of the photospheric absorption profile, enabling a robust decoupling of photospheric and HVF absorption in the \canir. We utilise a sample of 58 \sneia\ with well-sampled light curves and spectra within 5 days of maximum light from the Berkeley SN~Ia Program \citep[BSNIP;][]{bsnip1} sample to show that HVFs are stronger in more-slowly declining (i.e., low \dmft) \sneia\ and are generally absent in HV \sneia.

In Section~\ref{sec:snia_sample} we catalog the photometric and spectroscopic data in our sample, and describe the procedure for measuring the strength of HVFs in the spectra. Section~\ref{sec:hvfs_and_lcs} compares the measured HVF strengths to the light-curve properties and maximum-light \vsi\ values of our SN sample. In Section~\ref{sec:hvf_evolution}, the HVF evolution for a select few \sneia\ having extensive spectroscopic observations before maximum light is examined. We present concluding remarks in Section~\ref{sec:conclusions}.

\section{Dataset}
\label{sec:snia_sample}
\subsection{Spectroscopic and Photometric Samples}
\label{sec:samples}
The primary focus of this work is to measure HVFs in \snia\ spectra at maximum light, taking advantage of the fact that \siline\ HVFs have faded by maximum light but \canir\ HVFs persist for most \sneia. This task requires a sample of \sneia\ having well-sampled light curves so that the date of maximum light and the decline rate \dmft\ can be accurately measured. More importantly, we require a spectroscopic sample which extends far enough into the NIR to cover the \canir\ feature. The BSNIP \citep{bsnip1} sample -- collected primarily with the Kast double spectrograph \citep{kast} on the Shane 3\,m telescope at Lick Observatory -- is well suited to this study, as it includes a large number of objects with both photometric and spectroscopic time series, with the spectra covering a broad range of wavelength from the UV to the near IR.

In total, we found 61 \sneia\ from the BSNIP sample having spectra within 5 days of maximum light and well-sampled light curves. Three of these have insufficient signal-to-noise ratio (S/N) in the red to reliably measure the \canir\ absorption profile, leaving us a final sample of 58 \sneia. In Table~\ref{tab:lc_props} we report this list of \sneia\ in our dataset, the phase of the BSNIP spectrum used in the analysis, and the decline rate \dmft\ reported in the literature.

\begin{table}
\begin{center}
\caption{Light-Curve Properties and Spectrum Phases}
\label{tab:lc_props}
\begin{tabular}{lrrl}
\hline
SN & Phase  & $\Delta m_{15}(B)$ & LC Source$^a$\\
   & (days) & (mag)             & \\
\hline
SN1994S  &  0.8 & $1.05 \pm 0.04$ & CfA \\
SN1998es & -0.9 & $0.98 \pm 0.03$ & LOSS \\
SN1999aa & -0.5 & $0.94 \pm 0.01$ & CfA, LOSS \\
SN1999ac & -1.5 & $1.18 \pm 0.03$ & LOSS \\
SN2000cw &  4.7 & $1.31 \pm 0.04$ & LOSS \\
SN2000dk &  0.7 & $1.63 \pm 0.04$ & CfA, LOSS \\
SN2000dm & -2.4 & $1.56 \pm 0.05$ & LOSS \\
SN2000dn & -1.6 & $1.11 \pm 0.03$ & LOSS \\
SN2001br &  2.0 & $1.35 \pm 0.06$ & LOSS \\
SN2001cp &  0.6 & $0.93 \pm 0.04$ & CfA, LOSS \\
SN2001da & -1.6 & $1.25 \pm 0.05$ & CfA, LOSS \\
SN2001eh &  2.2 & $0.91 \pm 0.01$ & CfA, LOSS \\
SN2001ep &  2.5 & $1.34 \pm 0.02$ & CfA, LOSS \\
SN2001fe & -1.0 & $1.03 \pm 0.02$ & CfA \\
SN2002bo & -1.7 & $1.15 \pm 0.04$ & LOSS \\
SN2002cd &  0.3 & $0.96 \pm 0.03$ & LOSS \\
SN2002eb &  0.9 & $0.99 \pm 0.03$ & LOSS \\
SN2002ef &  4.0 & $1.04 \pm 0.10$ & LOSS \\
SN2002er & -4.2 & $1.28 \pm 0.05$ & LOSS \\
SN2002ha & -0.8 & $1.40 \pm 0.04$ & CfA, LOSS \\
SN2002he &  0.3 & $1.50 \pm 0.03$ & CfA, LOSS \\
SN2003cq & -3.0 & $1.26 \pm 0.05$ & CfA \\
SN2003he &  2.5 & $0.99 \pm 0.03$ & LOSS \\
SN2004gs &  0.1 & $1.60 \pm 0.02$ & CSP, LOSS \\
SN2005am &  3.6 & $1.48 \pm 0.03$ & CSP, LOSS \\
SN2005bc &  1.1 & $1.39 \pm 0.05$ & LOSS \\
SN2005cf & -1.8 & $1.08 \pm 0.03$ & LOSS \\
SN2005de & -1.3 & $1.22 \pm 0.03$ & LOSS \\
SN2005el &  0.8 & $1.36 \pm 0.01$ & CfA, CSP, LOSS \\
SN2005eq & -0.4 & $0.85 \pm 0.01$ & CfA, CSP, LOSS \\
SN2005ms & -2.6 & $1.10 \pm 0.01$ & CfA \\
SN2005na & -2.3 & $1.10 \pm 0.01$ & CfA, CSP, LOSS \\
SN2006D  &  1.6 & $1.39 \pm 0.02$ & CSP, LOSS \\
SN2006N  & -1.6 & $1.58 \pm 0.03$ & CfA \\
SN2006S  &  2.5 & $1.01 \pm 0.02$ & CfA \\
SN2006X  &  2.0 & $1.05 \pm 0.02$ & CSP, LOSS \\
SN2006bq &  4.1 & $1.45 \pm 0.03$ & CfA \\
SN2006bt &  2.4 & $1.03 \pm 0.01$ & CfA, CSP, LOSS \\
SN2006ef &  1.7 & $1.33 \pm 0.01$ & CfA, CSP, LOSS \\
SN2006ej & -4.0 & $1.35 \pm 0.01$ & CfA, CSP, LOSS \\
SN2006et &  3.2 & $1.09 \pm 0.01$ & CfA, CSP \\
SN2006gt &  0.9 & $1.66 \pm 0.03$ & CSP \\
SN2006kf & -3.0 & $1.48 \pm 0.01$ & CfA, CSP \\
SN2006sr &  1.6 & $1.39 \pm 0.04$ & CfA \\
SN2007A  & -0.2 & $0.95 \pm 0.02$ & CSP \\
SN2007F  &  2.6 & $1.04 \pm 0.01$ & CfA \\
SN2007S  &  3.9 & $1.06 \pm 0.01$ & CfA, CSP \\
SN2007af & -1.9 & $1.12 \pm 0.01$ & CSP, LOSS \\
SN2007ba &  4.5 & $1.89 \pm 0.03$ & CSP \\
SN2007bc & -0.0 & $1.23 \pm 0.01$ & CfA, CSP, LOSS \\
SN2007ci & -2.2 & $1.75 \pm 0.03$ & CfA, LOSS \\
SN2007co &  0.4 & $1.17 \pm 0.02$ & CfA, LOSS \\
SN2007fr & -0.5 & $1.79 \pm 0.04$ & LOSS \\
SN2007hj & -0.7 & $1.95 \pm 0.06$ & LOSS \\
SN2008ar &  2.8 & $1.08 \pm 0.05$ & LOSS \\
SN2008ec & -0.7 & $1.36 \pm 0.06$ & LOSS \\
\hline
\end{tabular}
\end{center}
$^a$ CfA = \citet{hicken09a}; CSP = \citet{contreras10}, \citet{stritzinger11}; LOSS = \citet{ganesh10}.
\end{table}

The light curves came primarily from three sources: the Harvard CfA SN group \citep{hicken09a}, the Carnegie Supernova Project \citep{contreras10, stritzinger11}, and the Lick Observatory Supernova Search \citep[LOSS;][]{ganesh10}.  While LOSS naturally had light-curve coverage for almost all the \sneia\ in our sample, the light curves from CfA or CSP were sometimes better sampled or had higher S/N. LOSS and CfA report their photometry on the standard $B$ photometric system, while CSP reports photometry in their instrument's natural system. \citet{ganesh11} found these systems to be consistent to within about 0.03~mag, and we found here that the decline rates were similarly consistent.

When a SN had multiple sources for its light curve, we calculate the final \dmft\ and its uncertainty as the weighted mean of values reported from all the groups. Although these three surveys used different light-curve fitters -- MLCS \citep{riess96, jha07} for CfA, SNooPy \citep{burns11} for CSP, and a mean light-curve template method for LOSS -- the value of \dmft\ measured from the fitted light curve provides a consistent measure of the light-curve decline rate across all methods. Generally we found the published \dmft\ values from different sources to agree within the quoted uncertainties.

To ensure that our sample does not suffer any major selection bias, we inspected the distribution of host-galaxy masses for those \sneia\ in our sample with host masses measured either by \citet{neill09} or \citet{kelly10}. We found that 35 of our 58 SNe had host masses from those studies, with a mean and $\pm1\sigma$ mass of $\log(M_*/{\rm M}_\odot) = 10.65 \pm 0.50$ for our host-mass distribution. This compares favourably to the values of $\log(M_*/{\rm M}_\odot) = 10.60 \pm 0.64$ for \citet{neill09} and $\log(M_*/{\rm M}_\odot) = 10.75 \pm 0.68$ for \citet{kelly10}. Thus, our sample has a host-mass distribution similar to that of the general sample of local \snia\ hosts. These \sneia\ arise from targeted surveys and have hosts with a higher average mass than untargeted surveys \citep[e.g., the SDSS-SN survey;][]{smith12}. However, the \dmft\ range of these \sneia\ spans the full range observed in untargeted surveys \citep[see, e.g.,][]{hicken09a}, so we have no reason to believe any significant biases exist in our sample.

\subsection{Measuring HVFs in \snia\ Spectra}
\label{sec:profile_fits}
Atomic transitions produce lines in \snia\ spectra which exhibit a P-Cygni profile as a result of the expanding ejecta. This profile consists of both emission and absorption components, and many overlapping lines form a ``pseudo-continuum.'' While the profiles of individual lines may have some strong influence from line emission \citep[see, e.g.,][]{vanrossum12}, analysis of the apparent absorption profiles of these lines remains an important empirical tool for analysing the composition of \snia\ ejecta. In the analysis below, we quantify the strength of HVFs by examining the normalised absorption profile in the \canir\ compared to the absorption profile of the \siline\ line. Our methods are based on, and closely resemble, those we employed in our analysis of SN~2012fr in \citet{childress12fr}.

\begin{figure*}
\begin{center}
\includegraphics[width=0.90\textwidth]{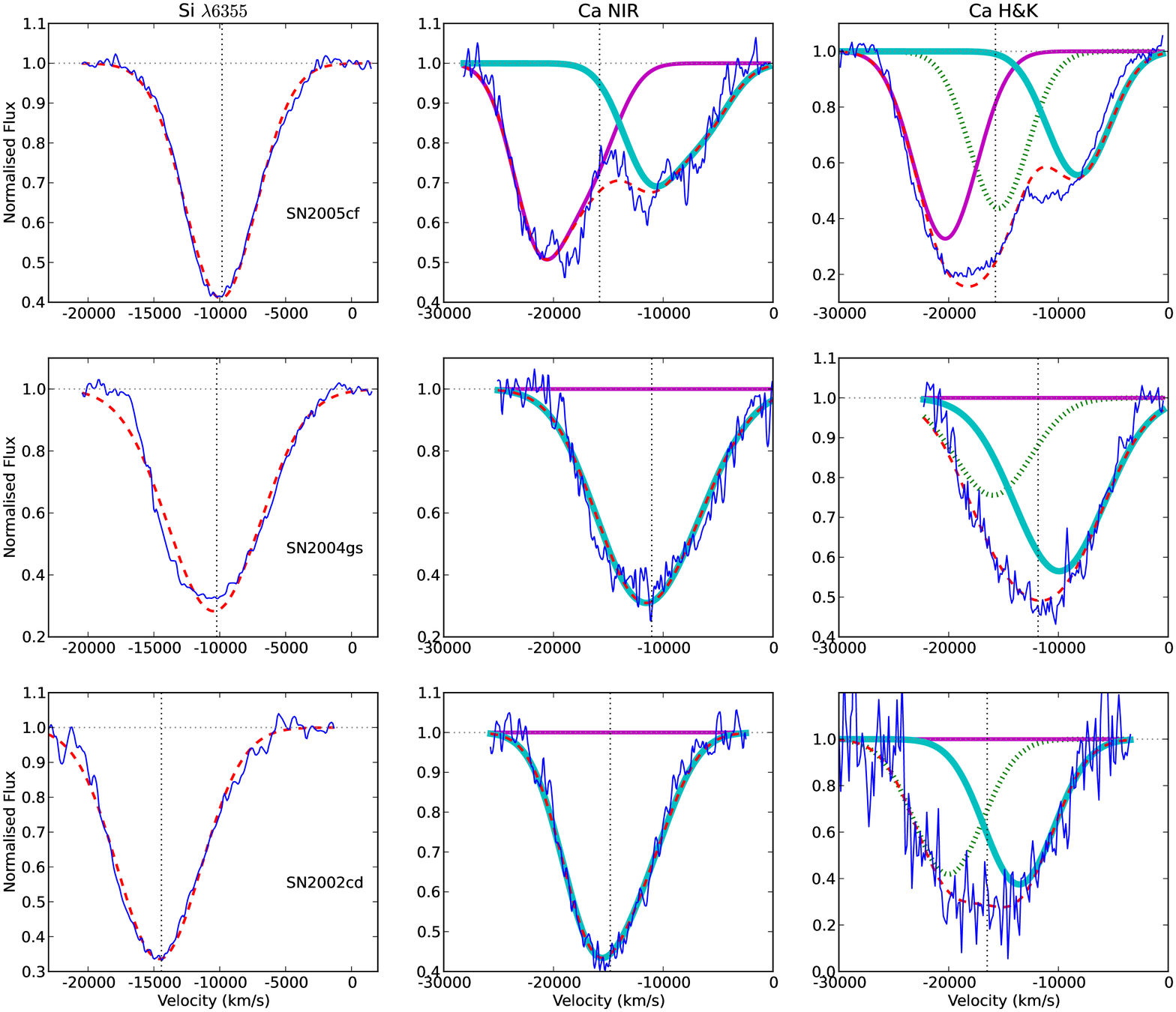}
\caption{Absorption-profile fits for the HVF-strong and slowly declining SN~2005cf (top row); the HVF-weak and rapidly declining SN~2004gs (middle row); and the HVF-weak, slowly declining, and high-velocity (HV) SN~2002cd (bottom row). Profiles have been normalised by the fitted pseudo-continuum, with data shown in blue and the full fitted profile in dashed red. For the \canir\ and \cahk\ profiles (middle and right columns, respectively), the fitted photospheric absorption component is indicated as the thin magenta curve, while the fitted HVF component is the thick cyan curve. For \cahk, the \siblue\ component is shown as the dotted green curve. The absorption-weighted velocity for each line is indicated as the vertical dotted black line in each panel (see text for details).}
\label{fig:hvf_profile_fits}
\end{center}
\end{figure*}

We begin the inspection of a given \snia\ absorption feature by first defining the pseudo-continuum of that feature. In practice, we assign (by visual inspection) regions of the spectrum to the blue and red of the feature of interest which are smooth and featureless. A pseudo-continuum is then defined for the full absorption-profile wavelength range by fitting a line to the flux in the blue and red pseudo-continuum regions. In each panel of Figure~\ref{fig:hvf_profile_fits}, the pseudo-continuum regions are shown as the shaded grey regions, and the fitted linear pseudo-continuum profile is shown as the dashed green line.

After defining the pseudo-continuum, we normalise the absorption profile by dividing the observed flux by the fitted pseudo-continuum profile. For the \siline\ line, we fit the normalised absorption profile as a single Gaussian profile in velocity space whose fit parameters are the velocity centre, velocity width, and peak absorption depth. We confirmed that the \siline\ profiles of all \sneia\ in this sample are well fit at maximum light by a single Gaussian, and show no evidence for HVFs in the \siline\ line.

The velocity centre, width, and absorption depth are fitted using a custom Python-based script which employs the {\tt mpfit} package to minimise the \chisq\ residuals between the observed flux and our simple parametrised line profile. To ensure that our \chisq\ values are meaningful, we derive flux errors for the BSNIP spectra as follows. We first smooth the SN spectrum with a Savitszky-Golay filter, using a third-order polynomial fit over a 50\,\AA\ range. We next determine the squared residuals of the data from the smoothed data, and then calculate a running mean at each wavelength bin using a window of 20\,\AA. This technique provides a smooth estimator of the flux error in the data, and produced favourable fits with $\chi^2_\nu$ values close to unity. The final fitted velocity centre, full width at half-maximum intensity (FWHM) velocity width, and integrated absorption area (i.e., the pseudo-equivalent width, pEW) are reported in Table~\ref{tab:prof_fit_results}. We also list the formal uncertainties on those quantities calculated from the covariance matrix returned by {\tt mpfit} from our fitting routine.

\begin{table*}
\begin{center}
\caption{Absorption-Profile Fit Results}
\label{tab:prof_fit_results}
\begin{tabular}{lrrrrrrrrr}
\hline
SN & \multicolumn{3}{c}{\siline} & \multicolumn{3}{c}{Photospheric Ca IR3} & \multicolumn{3}{c}{HVF Ca IR3} \\
& $v$    & FWHM & pEW   & $v$    & FWHM & pEW   & $v$    & FWHM & pEW \\
& (km\,s$^{-1}$) & (km\,s$^{-1}$) & (\AA) & (km\,s$^{-1}$) & (km\,s$^{-1}$) & (\AA) & (km\,s$^{-1}$) & (km\,s$^{-1}$) & (\AA) \\
\hline
SN~1994S  & $10190 \pm 10$ & $ 5650 \pm 20$ & $ 84 \pm  1$ & $10300 \pm  70$ & $ 5190 \pm  140$ & $109 \pm  5$ & $18120 \pm 190$ & $ 4470 \pm   10$ & $ 26 \pm  1$ \\
SN~1995D  & $ 9930 \pm 10$ & $ 5630 \pm 10$ & $ 82 \pm  1$ & $ 9750 \pm  20$ & $ 5060 \pm   10$ & $101 \pm  1$ & $16790 \pm  50$ & $ 5760 \pm  100$ & $ 57 \pm  2$ \\
SN~1995E  & $10520 \pm 10$ & $ 6750 \pm 10$ & $102 \pm  1$ & $10460 \pm  20$ & $ 6080 \pm   10$ & $105 \pm  1$ & $19060 \pm  60$ & $ 4860 \pm  120$ & $ 41 \pm  2$ \\
SN~1998es & $ 9930 \pm 10$ & $ 6100 \pm 20$ & $ 61 \pm  1$ & $ 9210 \pm  90$ & $ 5490 \pm   10$ & $ 32 \pm  1$ & $18410 \pm  40$ & $ 5430 \pm  100$ & $ 49 \pm  1$ \\
SN~1999aa & $10080 \pm 10$ & $ 5750 \pm 20$ & $ 58 \pm  1$ & $ 9820 \pm 120$ & $ 5180 \pm   10$ & $ 39 \pm  1$ & $18790 \pm 140$ & $ 5530 \pm  310$ & $ 40 \pm  3$ \\
SN~1999ac & $10110 \pm 10$ & $ 7510 \pm 20$ & $ 85 \pm  1$ & $10490 \pm  40$ & $ 6760 \pm   10$ & $ 81 \pm  1$ & $16950 \pm  20$ & $ 4470 \pm   10$ & $ 86 \pm  1$ \\
SN~2000cw & $ 9740 \pm 10$ & $ 6500 \pm 20$ & $112 \pm  1$ & $ 9550 \pm  10$ & $ 6630 \pm   10$ & $182 \pm  2$ & $16350 \pm  70$ & $ 5490 \pm  180$ & $ 98 \pm  5$ \\
SN~2000dk & $10440 \pm 10$ & $ 7370 \pm 10$ & $125 \pm  1$ & $11050 \pm  10$ & $ 6990 \pm   40$ & $263 \pm  3$ &         \nodata &          \nodata &          $0$ \\
SN~2000dm & $10740 \pm 10$ & $ 6510 \pm 10$ & $106 \pm  1$ & $ 9870 \pm  40$ & $ 5860 \pm   10$ & $109 \pm  1$ & $17990 \pm 120$ & $ 4600 \pm  240$ & $ 23 \pm  2$ \\
SN~2000dn & $10040 \pm 10$ & $ 6430 \pm 20$ & $101 \pm  1$ & $ 9550 \pm  50$ & $ 5790 \pm   10$ & $118 \pm  1$ & $17030 \pm 110$ & $ 4470 \pm  240$ & $ 47 \pm  4$ \\
SN~2001br & $12290 \pm 30$ & $ 8350 \pm 50$ & $ 90 \pm  1$ & $13060 \pm 100$ & $ 7520 \pm   10$ & $ 55 \pm  1$ &         \nodata &          \nodata &          $0$ \\
SN~2001cp & $10390 \pm 10$ & $ 5690 \pm 30$ & $ 81 \pm  1$ & $10190 \pm  90$ & $ 5120 \pm   10$ & $ 89 \pm  2$ & $17300 \pm 250$ & $ 5870 \pm  550$ & $ 39 \pm  5$ \\
SN~2001da & $11080 \pm 10$ & $ 7510 \pm 20$ & $117 \pm  1$ & $10590 \pm  50$ & $ 6760 \pm   10$ & $112 \pm  1$ & $18660 \pm  40$ & $ 5280 \pm   80$ & $105 \pm  2$ \\
SN~2001eh & $10610 \pm 10$ & $ 4870 \pm 20$ & $ 60 \pm  1$ & $10220 \pm  60$ & $ 4380 \pm   10$ & $ 64 \pm  1$ & $17870 \pm  70$ & $ 5120 \pm  170$ & $ 63 \pm  4$ \\
SN~2001ep & $ 9640 \pm 10$ & $ 7010 \pm 10$ & $117 \pm  1$ & $ 9200 \pm  20$ & $ 7720 \pm   10$ & $156 \pm  1$ & $15000 \pm  10$ & $ 7540 \pm  120$ & $ 97 \pm  2$ \\
SN~2001fe & $10720 \pm 10$ & $ 5670 \pm 10$ & $ 73 \pm  1$ & $11340 \pm  50$ & $ 5100 \pm   10$ & $ 48 \pm  1$ &         \nodata &          \nodata &          $0$ \\
SN~2002bo & $12860 \pm 10$ & $ 9090 \pm 30$ & $150 \pm  1$ & $13570 \pm  20$ & $ 8670 \pm   70$ & $217 \pm  3$ &         \nodata &          \nodata &          $0$ \\
SN~2002cd & $14720 \pm 10$ & $ 7350 \pm 20$ & $115 \pm  1$ & $14930 \pm  20$ & $ 6610 \pm   10$ & $156 \pm  1$ &         \nodata &          \nodata &          $0$ \\
SN~2002eb & $ 9960 \pm 10$ & $ 5450 \pm 20$ & $ 75 \pm  1$ & $ 8970 \pm  10$ & $ 4900 \pm   10$ & $ 56 \pm  1$ & $18560 \pm  50$ & $ 4980 \pm  130$ & $ 53 \pm  3$ \\
SN~2002ef & $11110 \pm 10$ & $ 6130 \pm 30$ & $106 \pm  1$ & $10220 \pm 170$ & $ 5560 \pm  260$ & $139 \pm  8$ & $15670 \pm 470$ & $ 5340 \pm  650$ & $ 50 \pm  7$ \\
SN~2002er & $11730 \pm 10$ & $ 6860 \pm 10$ & $107 \pm  1$ & $10950 \pm  20$ & $ 6280 \pm   50$ & $121 \pm  1$ & $19440 \pm  20$ & $ 5620 \pm   40$ & $ 93 \pm  1$ \\
SN~2002ha & $10680 \pm 10$ & $ 7020 \pm 10$ & $108 \pm  1$ & $10330 \pm  30$ & $ 6320 \pm   10$ & $113 \pm  1$ & $17350 \pm  90$ & $ 4470 \pm   10$ & $ 29 \pm  1$ \\
SN~2002he & $12020 \pm 10$ & $ 7420 \pm 10$ & $127 \pm  1$ & $11880 \pm  30$ & $ 7340 \pm   80$ & $175 \pm  3$ & $18830 \pm 330$ & $ 4470 \pm   10$ & $  7 \pm  1$ \\
SN~2003cq & $11750 \pm 20$ & $ 7520 \pm 40$ & $127 \pm  1$ & $12000 \pm  50$ & $ 8270 \pm   10$ & $184 \pm  1$ &         \nodata &          \nodata &          $0$ \\
SN~2003he & $11040 \pm 10$ & $ 5900 \pm 20$ & $104 \pm  1$ & $10550 \pm  80$ & $ 5310 \pm   10$ & $145 \pm  3$ & $17010 \pm 210$ & $ 5370 \pm  410$ & $ 54 \pm  6$ \\
SN~2004gs & $10480 \pm 10$ & $ 8160 \pm 20$ & $135 \pm  1$ & $11210 \pm  20$ & $ 8970 \pm   10$ & $234 \pm  1$ &         \nodata &          \nodata &          $0$ \\
SN~2005am & $10870 \pm 10$ & $ 7260 \pm 10$ & $118 \pm  1$ & $11100 \pm  10$ & $ 7980 \pm   10$ & $202 \pm  1$ & $20000 \pm  10$ & $ 6710 \pm   10$ & $  1 \pm  1$ \\
SN~2005bc & $10430 \pm 10$ & $ 6960 \pm 10$ & $116 \pm  1$ & $10730 \pm  30$ & $ 7650 \pm   10$ & $182 \pm  1$ & $17260 \pm 120$ & $ 5160 \pm  210$ & $ 29 \pm  2$ \\
SN~2005cf & $ 9900 \pm 10$ & $ 6120 \pm 10$ & $ 86 \pm  1$ & $ 9520 \pm  50$ & $ 5510 \pm   10$ & $ 76 \pm  1$ & $19670 \pm  20$ & $ 5350 \pm   70$ & $119 \pm  3$ \\
SN~2005de & $10110 \pm 10$ & $ 6460 \pm 30$ & $102 \pm  1$ & $10110 \pm  50$ & $ 5810 \pm   10$ & $127 \pm  1$ & $18120 \pm 110$ & $ 4640 \pm  220$ & $ 50 \pm  4$ \\
SN~2005el & $10190 \pm 10$ & $ 6310 \pm 20$ & $ 93 \pm  1$ & $ 9730 \pm  30$ & $ 5680 \pm   10$ & $101 \pm  1$ & $17370 \pm  90$ & $ 4470 \pm   10$ & $ 32 \pm  1$ \\
SN~2005eq & $ 9730 \pm 10$ & $ 5430 \pm 20$ & $ 55 \pm  1$ & $ 8880 \pm 150$ & $ 4880 \pm   10$ & $ 30 \pm  1$ & $18570 \pm  80$ & $ 5330 \pm  200$ & $ 46 \pm  3$ \\
SN~2005ms & $11480 \pm 20$ & $ 7490 \pm 40$ & $119 \pm  1$ & $10650 \pm 280$ & $ 6850 \pm  640$ & $ 98 \pm 14$ & $18550 \pm 220$ & $ 4470 \pm   10$ & $ 57 \pm  5$ \\
SN~2005na & $10220 \pm 10$ & $ 5490 \pm 40$ & $ 70 \pm  1$ & $ 9830 \pm 130$ & $ 4940 \pm   10$ & $ 44 \pm  1$ &         \nodata &          \nodata &          $0$ \\
SN~2006D  & $10210 \pm 10$ & $ 6390 \pm 10$ & $100 \pm  1$ & $11750 \pm  10$ & $ 7040 \pm   20$ & $196 \pm  1$ &         \nodata &          \nodata &          $0$ \\
SN~2006N  & $11110 \pm 10$ & $ 6910 \pm 20$ & $114 \pm  1$ & $10600 \pm  50$ & $ 6220 \pm   10$ & $138 \pm  1$ & $17800 \pm 230$ & $ 4470 \pm   10$ & $ 22 \pm  1$ \\
SN~2006S  & $10430 \pm 10$ & $ 5170 \pm 30$ & $ 73 \pm  1$ & $ 9780 \pm 110$ & $ 4650 \pm   10$ & $ 82 \pm  3$ & $17930 \pm 180$ & $ 6060 \pm  420$ & $ 82 \pm  9$ \\
SN~2006X  & $14890 \pm 10$ & $ 9750 \pm 10$ & $186 \pm  1$ & $15430 \pm  10$ & $10640 \pm   10$ & $320 \pm  1$ &         \nodata &          \nodata &          $0$ \\
SN~2006bq & $12910 \pm 10$ & $ 9010 \pm 10$ & $166 \pm  1$ & $12510 \pm  20$ & $ 9910 \pm   10$ & $267 \pm  1$ & $20000 \pm  10$ & $ 2500 \pm   90$ & $ 16 \pm  1$ \\
SN~2006bt & $10310 \pm 10$ & $ 8200 \pm 20$ & $130 \pm  1$ & $ 9280 \pm  10$ & $ 8680 \pm  160$ & $163 \pm  5$ & $16740 \pm  50$ & $ 5710 \pm  130$ & $103 \pm  4$ \\
SN~2006ef & $11610 \pm 10$ & $ 7820 \pm 10$ & $134 \pm  1$ & $10800 \pm  20$ & $ 7040 \pm   10$ & $169 \pm  1$ & $20270 \pm 140$ & $ 5220 \pm  350$ & $ 20 \pm  2$ \\
SN~2006ej & $12210 \pm 10$ & $ 7690 \pm 10$ & $118 \pm  1$ & $11880 \pm  30$ & $ 6920 \pm   10$ & $123 \pm  1$ & $18700 \pm 130$ & $ 4470 \pm   10$ & $ 15 \pm  1$ \\
SN~2006et & $ 9730 \pm 10$ & $ 5910 \pm 20$ & $ 74 \pm  1$ & $ 9300 \pm  80$ & $ 5320 \pm   10$ & $ 55 \pm  1$ & $17500 \pm  40$ & $ 5570 \pm  110$ & $ 94 \pm  3$ \\
SN~2006gt & $ 9660 \pm 20$ & $ 6910 \pm 40$ & $118 \pm  1$ & $10630 \pm  10$ & $ 7600 \pm   10$ & $228 \pm  1$ &         \nodata &          \nodata &          $0$ \\
SN~2006kf & $11070 \pm 10$ & $ 7400 \pm 10$ & $116 \pm  1$ & $ 9960 \pm  10$ & $ 6660 \pm   10$ & $100 \pm  1$ & $16960 \pm 110$ & $ 4470 \pm   10$ & $ 15 \pm  1$ \\
SN~2006sr & $11450 \pm 10$ & $ 7170 \pm 10$ & $116 \pm  1$ & $11130 \pm  50$ & $ 6820 \pm   90$ & $157 \pm  3$ & $18040 \pm 320$ & $ 4650 \pm  520$ & $ 12 \pm  2$ \\
SN~2007A  & $10420 \pm 10$ & $ 5970 \pm 10$ & $ 88 \pm  1$ & $ 9850 \pm  30$ & $ 5370 \pm   10$ & $ 83 \pm  1$ & $16810 \pm  60$ & $ 4470 \pm   10$ & $ 25 \pm  1$ \\
SN~2007F  & $10460 \pm 10$ & $ 5450 \pm 10$ & $ 86 \pm  1$ & $10720 \pm  40$ & $ 4900 \pm   10$ & $ 97 \pm  1$ & $17010 \pm 180$ & $ 4470 \pm   10$ & $ 21 \pm  1$ \\
SN~2007S  & $10120 \pm 10$ & $ 5790 \pm 20$ & $ 66 \pm  1$ & $10000 \pm  50$ & $ 5210 \pm   10$ & $ 67 \pm  1$ & $17260 \pm  50$ & $ 5260 \pm   90$ & $ 75 \pm  2$ \\
SN~2007af & $10430 \pm 10$ & $ 6650 \pm 10$ & $106 \pm  1$ & $10370 \pm  20$ & $ 6230 \pm   40$ & $142 \pm  1$ & $17920 \pm  60$ & $ 4470 \pm   10$ & $ 32 \pm  1$ \\
SN~2007ba & $ 9550 \pm 20$ & $ 7410 \pm 50$ & $111 \pm  1$ & $10430 \pm  30$ & $ 6890 \pm  110$ & $255 \pm  7$ &         \nodata &          \nodata &          $0$ \\
SN~2007bc & $ 9680 \pm 10$ & $ 6520 \pm 10$ & $103 \pm  1$ & $ 9510 \pm  20$ & $ 5870 \pm   10$ & $135 \pm  1$ & $16670 \pm  60$ & $ 4060 \pm  110$ & $ 38 \pm  2$ \\
SN~2007ci & $11520 \pm 10$ & $ 7250 \pm 20$ & $122 \pm  1$ & $11580 \pm  50$ & $ 6520 \pm   10$ & $138 \pm  1$ & $17760 \pm 1360$ & $ 4470 \pm   10$ & $  3 \pm  2$ \\
SN~2007co & $11210 \pm 10$ & $ 7270 \pm 20$ & $120 \pm  1$ & $10620 \pm  40$ & $ 6550 \pm   10$ & $138 \pm  1$ & $17530 \pm  40$ & $ 4900 \pm   80$ & $ 98 \pm  2$ \\
SN~2007fr & $10440 \pm 20$ & $ 6870 \pm 50$ & $112 \pm  1$ & $10810 \pm 200$ & $ 6230 \pm  350$ & $232 \pm 14$ &         \nodata &          \nodata &          $0$ \\
SN~2007hj & $11500 \pm 10$ & $ 8490 \pm 10$ & $147 \pm  1$ & $12130 \pm  10$ & $ 9340 \pm   10$ & $264 \pm  1$ & $20000 \pm  10$ & $ 4470 \pm   10$ & $ 16 \pm  1$ \\
SN~2008ar & $ 9990 \pm 20$ & $ 5980 \pm 50$ & $ 91 \pm  1$ & $ 8990 \pm  10$ & $ 5580 \pm  420$ & $ 92 \pm 12$ & $16870 \pm 180$ & $ 4470 \pm   10$ & $ 65 \pm  3$ \\
SN~2008ec & $10410 \pm 10$ & $ 7140 \pm 10$ & $122 \pm  1$ & $ 9820 \pm  30$ & $ 6430 \pm   10$ & $127 \pm  1$ & $16580 \pm 140$ & $ 4470 \pm   10$ & $ 17 \pm  1$ \\
\hline
\end{tabular}
\end{center}
\end{table*}

Fitting the \canir\ absorption profile is more complicated than the \siline\ line. As a multiplet, the absorption profile in wavelength space for a single absorbing system in velocity space is the sum of absorption profiles for the three lines weighted by their relative strength. In the optically thin regime, these would be the Einstein $B$ values. In the optically thick regime more typical of SN atmospheres, the lines would tend to saturate and thus have absorption strengths approaching equality \citep[see the discussion in][for more details]{childress12fr}. For our fits, we assume the optically thick regime and fit for two components in velocity space. We force the velocity centre and velocity width of the photospheric component of the \canir\ to be within 10\% of the values measured for the \siline\ line. The HVF component of the \canir\ feature is effectively unconstrained other than the velocity centre being forced to be higher than that of the photospheric component by 2000 \kms. Constraining the HVF velocity centre to be higher than the photospheric component was required in order to prevent the fitting routine from finding the same centre for both components, and we determined that 2000 \kms\ achieved successful separation of the two components while preventing the fitted HVF centre from occurring at its lower bound. The \canir\ fit results are reported in Table~\ref{tab:prof_fit_results}.

In the analysis that follows, we will parametrise the strength of the HVFs by the ratio of the HVF absorption to the photospheric absorption (i.e., the pEWs).  Specifically, we will call this ratio \rhvf\ as defined by
\begin{equation}
  R_{\rm HVF} = \frac{{\rm pEW}({\rm HVF}_{\rm CaIR})}{{\rm pEW}({\rm Phot}_{\rm CaIR})},
\end{equation}
where the pEW values are calculated from the absorption-profile fits as described above.

For elements of discussion below, we are also interested in the impact of HVFs on the \cahk\ absorption feature. As noted by previous authors \citep{maguire12, foley12}, this line complex is affected by the \siblue\ line, complicating the interpretation of HVFs in the line. As in \citet{childress12fr}, we fit this line complex with three components: photospheric \cahk, HVF \cahk, and the \siblue\ line. In this work, we fixed the velocity centre and widths of these three components from the photospheric \canir, HVF \canir, and \siline\ line-fit results, respectively, and fit only for the relative absorption strengths. Most cases resulted in the HVF \cahk\ and \siblue\ line being degenerate in wavelength space, making it impossible to decompose the impact of these two components (\siblue\ corresponds to \cahk\ at $v \approx 5700$~\kms, so this degeneracy is strongest when the velocity difference between photospheric and HVF Ca is in this range). Only in rare cases where the HVF \cahk\ was at extremely high velocity (such as SN~2005cf, as shown in Figure~\ref{fig:hvf_profile_fits}) could clean separation of these two components be successfully achieved. This underscores the importance of having the clean \canir\ absorption feature to effectively measure the velocity structure of the HVFs.

Because of this degeneracy, we do not report the results of those fits in Table~\ref{tab:prof_fit_results}, but we do give in Table~\ref{tab:no_hvf_cahk} the pEW decomposition of \cahk\ and \siblue\ for cases without HVFs. In this table we also report the ratio of pEW values for the \siblue\ line to both \cahk\ and to \siline, with \sneia\ ordered by their \dmft\ values. With these data we can comment on some of the trends predicted by \citet{foley12}. First, we find that nearly half our sample is free of HVFs, roughly consistent with the prediction by \citet{foley12} that most \sneia\ should have the \cahk\ complex free of HVFs. We also find that $\sim 48$\% of the absorption strength in the \cahk\ line complex appears to be caused by \siblue\ in the HVF-free sample. Finally, we find no strong evidence for a trend of the \siblue\ strength with \dmft, either when comparing to the true \cahk\ absorption or to the \siline\ absorption strength.

\begin{table}
\begin{center}
\caption{\SiII\ and \cahk\ in no-HVF \sneia, ordered by \dmft}
\label{tab:no_hvf_cahk}
\begin{tabular}{lrrrrr}
\hline
SN & pEW(Si) & pEW(Ca) & $R_{CS}$~$^\mathrm{a}$ & $R_{Si}$~$^\mathrm{b}$ & \dmft\ \\
   & (\AA)   & (\AA)   &                    &                   & (mag)  \\
\hline
SN~2002cd & $ 61$ & $ 66$ & $0.48$ & $0.53$ & $0.96$  \\
SN~2006bt & $ 61$ & $ 48$ & $0.56$ & $0.47$ & $1.03$  \\
SN~2006X  & $ 65$ & $ 91$ & $0.42$ & $0.35$ & $1.05$  \\
SN~2005na & $ 33$ & $ 50$ & $0.40$ & $0.48$ & $1.10$  \\
SN~2007af & $ 62$ & $ 58$ & $0.51$ & $0.58$ & $1.12$  \\
SN~2002bo & $ 70$ & $ 67$ & $0.51$ & $0.47$ & $1.15$  \\
SN~2007co & $ 95$ & $ 49$ & $0.66$ & $0.79$ & $1.17$  \\
SN~1999ac & $ 67$ & $ 49$ & $0.58$ & $0.78$ & $1.18$  \\
SN~2007bc & $ 59$ & $ 53$ & $0.53$ & $0.57$ & $1.23$  \\
SN~2003cq & $ 76$ & $ 64$ & $0.54$ & $0.60$ & $1.26$  \\
SN~2006ej & $ 45$ & $ 48$ & $0.48$ & $0.38$ & $1.35$  \\
SN~2006D  & $ 29$ & $ 74$ & $0.28$ & $0.29$ & $1.39$  \\
SN~2006sr & $ 53$ & $ 54$ & $0.49$ & $0.46$ & $1.39$  \\
SN~2006bq & $ 68$ & $ 63$ & $0.52$ & $0.41$ & $1.45$  \\
SN~2005am & $ 40$ & $ 49$ & $0.45$ & $0.34$ & $1.48$  \\
SN~2006kf & $ 67$ & $ 36$ & $0.65$ & $0.58$ & $1.49$  \\
SN~2002he & $ 58$ & $ 61$ & $0.49$ & $0.45$ & $1.50$  \\
SN~2000dm & $ 51$ & $ 48$ & $0.52$ & $0.48$ & $1.56$  \\
SN~2006N  & $ 52$ & $ 46$ & $0.53$ & $0.46$ & $1.57$  \\
SN~2004gs & $ 29$ & $ 59$ & $0.33$ & $0.21$ & $1.60$  \\
SN~2000dk & $ 57$ & $ 61$ & $0.48$ & $0.45$ & $1.63$  \\
SN~2006gt & $ 32$ & $ 73$ & $0.31$ & $0.27$ & $1.66$  \\
SN~2007fr & $ 50$ & $ 56$ & $0.47$ & $0.45$ & $1.79$  \\
SN~2007ba & $ 46$ & $ 70$ & $0.40$ & $0.41$ & $1.89$  \\
SN~2007hj & $ 46$ & $ 70$ & $0.40$ & $0.32$ & $1.95$  \\
\hline
\end{tabular}
$^\mathrm{a}$ $R_{CS} = $ pEW(\siblue) / pEW(\cahk) \\
$^\mathrm{b}$ $R_{Si} = $ pEW(\siblue) / pEW(\siline)
\end{center}
\end{table}

We now briefly comment on the possible effect of our choice to constrain the photospheric \canir\ velocity profile to be close to that of the \siline\ line. For \sneia\ where the HVF in the \canir\ is at much higher velocity (such as SN~2005cf, SN~2009ig, and SN~2012fr; see Section~\ref{sec:hvf_evolution}), we found that an unconstrained fit yielded the same velocity profile as when the photospheric velocity centre and width were constrained by the \siline\ results. These same analyses generally found the \canir\ photospheric velocity to be consistent with that of the \siline\ velocity to within roughly 10\%, so this value was used to set the velocity constraints. The difficulty arises when the HVF feature has a velocity only a few thousand \kms\ higher than that of the photospheric component. In these cases an unconstrained fit often yielded very narrow photospheric components and very broad HVF components, since the two components overlap in velocity space. The only way to break this degeneracy in velocity space was to impose the independent constraint obtained from the \siline\ measurements. We acknowledge that the pEW decomposition of the \canir\ profile may be imperfect owing to these constraints, but forcing a consistency between \CaII\ and \SiII\ seems reasonable since we found that the velocities of these two species tend to be consistent in cases where the photospheric and HVF components cleanly separate or in cases where the lines are entirely photospheric (i.e., free of HVFs).

To provide a cross-check to our absorption-profile fits, we also desired a more model-independent measurement of the velocity profile of the absorption features. To this end, we calculated the weighted mean absorption velocity for each line as
\begin{equation}
  \bar{v}_X = \frac{\int v\,\cdot a(v) ~dv}{\int a(v)~dv},
\end{equation}
where $a(v)$ is the normalised absorption profile (flux divided by pseudo-continuum) in velocity space. In practice, this is calculated as the normalised absorption profile in wavelength space transformed to velocity space using the mean rest wavelength of the absorption line. For \siline\ this is simply $\lambda=6355$~\AA, and for the \CaII\ multiplets we assume the optically thick regime which results in simply the mean wavelength of all the lines in the multiplet ($\lambda=8567$~\AA\ for the \canir, and $\lambda=3951$~\AA\ for \cahk). These absorption-weighted velocities are reported in Table~\ref{tab:abs_vels}. We denote the weighted mean absorption velocities for \siline, the \canir, and \cahk\ as \vbsi, \vbci, and \vbch, respectively.

\begin{table}
\begin{center}
\caption{Weighted Absorption Velocities and HVF Ratio}
\label{tab:abs_vels}
\begin{tabular}{lrrrr}
\hline
SN & \vbsi\ & \vbci\ & \vbch & \rhvf\ \\
   & (km\,s$^{-1}$) & (km\,s$^{-1}$) & (km\,s$^{-1}$) & \\
\hline
SN~1994S  &   9940 & 12110 & 11620 & 0.24 +- 0.01 \\
SN~1995D  &   9770 & 12440 & 10930 & 0.57 +- 0.02 \\
SN~1995E  &  10420 & 12910 & 12090 & 0.39 +- 0.02 \\
SN~1998es &   9950 & 15180 & 11580 & 1.50 +- 0.04 \\
SN~1999aa &   9970 & 14720 & 10250 & 1.03 +- 0.08 \\
SN~1999ac &   9920 & 14020 & 11280 & 1.06 +- 0.00 \\
SN~2000cw &   9600 & 11860 & 12710 & 0.54 +- 0.03 \\
SN~2000dk &  10190 & 11470 & 12050 & 0.00 +- 0.00 \\
SN~2000dm &  10540 & 11580 & 11960 & 0.21 +- 0.02 \\
SN~2000dn &   9970 & 11680 & 10790 & 0.40 +- 0.04 \\
SN~2001br &  12010 & 13070 & 12340 & 0.00 +- 0.00 \\
SN~2001cp &  10220 & 12500 & 11620 & 0.44 +- 0.06 \\
SN~2001da &  10960 & 14760 & 14940 & 0.94 +- 0.02 \\
SN~2001eh &  10570 & 13890 & 13050 & 0.98 +- 0.05 \\
SN~2001ep &   9370 & 11530 & 11980 & 0.63 +- 0.01 \\
SN~2001fe &  10470 & 11040 & 10980 & 0.00 +- 0.00 \\
SN~2002bo &  12500 & 13520 & 14030 & 0.00 +- 0.00 \\
SN~2002cd &  14420 & 14830 & 15120 & 0.00 +- 0.00 \\
SN~2002eb &   9810 & 14000 & 11950 & 0.95 +- 0.05 \\
SN~2002ef &  10930 & 11710 & 11610 & 0.36 +- 0.05 \\
SN~2002er &  11570 & 14870 & 14790 & 0.77 +- 0.01 \\
SN~2002ha &  10560 & 12050 & 11670 & 0.25 +- 0.01 \\
SN~2002he &  11670 & 12380 & 13010 & 0.04 +- 0.01 \\
SN~2003cq &  11310 & 12120 & 13950 & 0.00 +- 0.00 \\
SN~2003he &  10890 & 12610 & 13080 & 0.37 +- 0.04 \\
SN~2004gs &  10230 & 11040 & 10480 & 0.00 +- 0.00 \\
SN~2005am &  10760 & 11280 & 12890 & 0.00 +- 0.00 \\
SN~2005bc &  10160 & 11830 & 10820 & 0.16 +- 0.01 \\
SN~2005cf &   9830 & 15800 & 14360 & 1.57 +- 0.04 \\
SN~2005de &   9900 & 12620 & 13060 & 0.40 +- 0.03 \\
SN~2005el &   9980 & 11970 & 11720 & 0.32 +- 0.00 \\
SN~2005eq &   9720 & 14610 & 11550 & 1.52 +- 0.10 \\
SN~2005ms &  11310 & 13670 & 13010 & 0.58 +- 0.05 \\
SN~2005na &   9960 &  9700 & 10090 & 0.00 +- 0.00 \\
SN~2006D  &  10050 & 12030 &  9960 & 0.00 +- 0.00 \\
SN~2006N  &  10830 & 11720 & 12530 & 0.16 +- 0.01 \\
SN~2006S  &  10330 & 14270 & 12790 & 1.00 +- 0.11 \\
SN~2006X  &  14380 & 15170 & 15360 & 0.00 +- 0.00 \\
SN~2006bq &  12640 & 12920 & 14020 & 0.06 +- 0.00 \\
SN~2006bt &   9980 & 12570 & 11570 & 0.64 +- 0.03 \\
SN~2006ef &  11320 & 12000 & 13390 & 0.12 +- 0.01 \\
SN~2006ej &  11830 & 12690 & 12420 & 0.12 +- 0.00 \\
SN~2006et &   9830 & 14120 & 12260 & 1.69 +- 0.05 \\
SN~2006gt &   9430 & 11360 &  9610 & 0.00 +- 0.00 \\
SN~2006kf &  11040 & 11170 & 13300 & 0.15 +- 0.00 \\
SN~2006sr &  11230 & 11760 & 12080 & 0.08 +- 0.01 \\
SN~2007A  &  10260 & 11610 & 10670 & 0.30 +- 0.00 \\
SN~2007F  &  10280 & 12000 & 11040 & 0.22 +- 0.01 \\
SN~2007S  &  10100 & 13930 & 12310 & 1.11 +- 0.03 \\
SN~2007af &  10070 & 11970 & 12170 & 0.23 +- 0.00 \\
SN~2007ba &   9210 & 11700 &  9510 & 0.01 +- 0.02 \\
SN~2007bc &   9480 & 11410 & 11540 & 0.28 +- 0.01 \\
SN~2007ci &  11230 & 11900 & 12010 & 0.02 +- 0.01 \\
SN~2007co &  11080 & 13520 & 13930 & 0.71 +- 0.02 \\
SN~2007fr &  10260 & 12600 & 10860 & 0.01 +- 0.04 \\
SN~2007hj &  11060 & 12590 & 11910 & 0.06 +- 0.00 \\
SN~2008ar &   9820 & 12540 & 13720 & 0.70 +- 0.03 \\
SN~2008ec &  10060 & 10970 & 11500 & 0.13 +- 0.01 \\
\hline
\end{tabular}
\end{center}
\end{table}


Finally, we revisit the issue of uncertainties in the main quantities derived in our analysis. As a representative case we discuss the uncertainty on \vsi, which is typically the least uncertain quantity we measure. The formal uncertainty, as propagated from flux errors, is typically of order 10s to 100 \kms. This formal error fails to capture the uncertainty caused by non-Gaussianity of the absorption profile, and uncertainty in the true flux level of the pseudo-continuum caused by the presence of neighbouring lines. These are general unsolved problems in absorption-feature fitting for all types of SNe, and common practice is to assume the uncertainty in the absorption velocity to be a few hundred \kms, which is roughly 5\% of the velocity width of the absorption profile. Thus, we consider a value of 300~\kms\ to be a reasonable estimate of the {\em systematic} uncertainty in both the fitted velocity centres of our absorption profiles and the absorption-weighted mean velocities (which suffer from the same pseudo-continuum uncertainties).

Uncertainty in the relative HVF strength (\rhvf) is more difficult to quantify. Because the photospheric and HVF components frequently have nonzero overlap in velocity (and wavelength) space, the dominant source of error here would be an error in measurement of the central velocity and velocity width of the photospheric component. We found above that in cases with cleanly separable photospheric components of \siline\ and the \canir, these lines had consistent velocities within about 1000~\kms\ (which is roughly consistent with the velocity constraint placed on this line by our requirement that the photospheric \canir\ be within 10\% of the velocity of \siline, which averages about 10,000~\kms). The widths of the photospheric \canir\ components were typically in the range 5000--8000~\kms, meaning that an error of 1000~\kms\ in the centre equates to perhaps 15--20\% uncertainty in the pEW of that component. Since most \rhvf\ values are less than 1, this is the dominant source of error, so uncertainties in \rhvf\ should be of order 15--20\% or less. Since our primary qualitative results will be demonstrated robustly with both \rhvf\ and the absorption-weighted velocity differences, profile-fitting uncertainties should not significantly affect our conclusions.

\section{HVFs: Relationships to \snia\ Light Curves and Velocities}
\label{sec:hvfs_and_lcs}
\subsection{HVFs and Light-Curve Decline Rate}
\label{sec:hvf_vs_dmft}
\begin{figure}
\begin{center}
\includegraphics[width=0.45\textwidth]{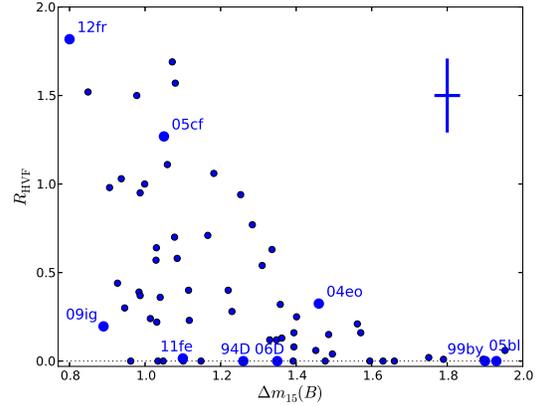}
\caption{Ratio of absorption pseudo-equivalent width (pEW) values for the photospheric and HVF components of the \canir\ (following the procedure outlined in Section~\ref{sec:profile_fits}) plotted against the light-curve decline rate \dmft\ (compiled in Section~\ref{sec:samples} and Table~\ref{tab:lc_props}). \sneia\ with spectral time series analysed in Section~\ref{sec:hvf_evolution} are labeled by name and have larger symbols.}
\label{fig:hvf_vs_dm15}
\end{center}
\end{figure}

Now that we have quantified the strength of the HVFs in our \snia\ sample, we inspect how the HVF strengths vary with other observable properties of the SNe. We begin with the light-curve decline rate \dmft, and show in Figure~\ref{fig:hvf_vs_dm15} the HVF pEW ratio \rhvf\ plotted versus \dmft\ for our sample. The most apparent result of this analysis is that very rapidly declining ($\Delta m_{15}(B) \geq 1.4$ mag) \sneia\ have weak or no HVFs whatsoever. An example from this class, SN~2004gs ($\Delta m_{15}(B) = 1.60$ mag), is shown in the middle panels of Figure~\ref{fig:hvf_profile_fits}, and clearly demonstrates a similar velocity profile for the \canir\ as for the \siline\ line. Figure~\ref{fig:hvf_vs_dm15} also indicates that HVF strength tends to increase, on average, as the light-curve decline rate decreases (i.e., for the brighter \sneia), as can be seen for SN~2005cf ($\Delta m_{15}(B) = 1.08$ mag, shown in the top panels of Figure~\ref{fig:hvf_profile_fits}).

\begin{figure}
\begin{center}
\includegraphics[width=0.45\textwidth]{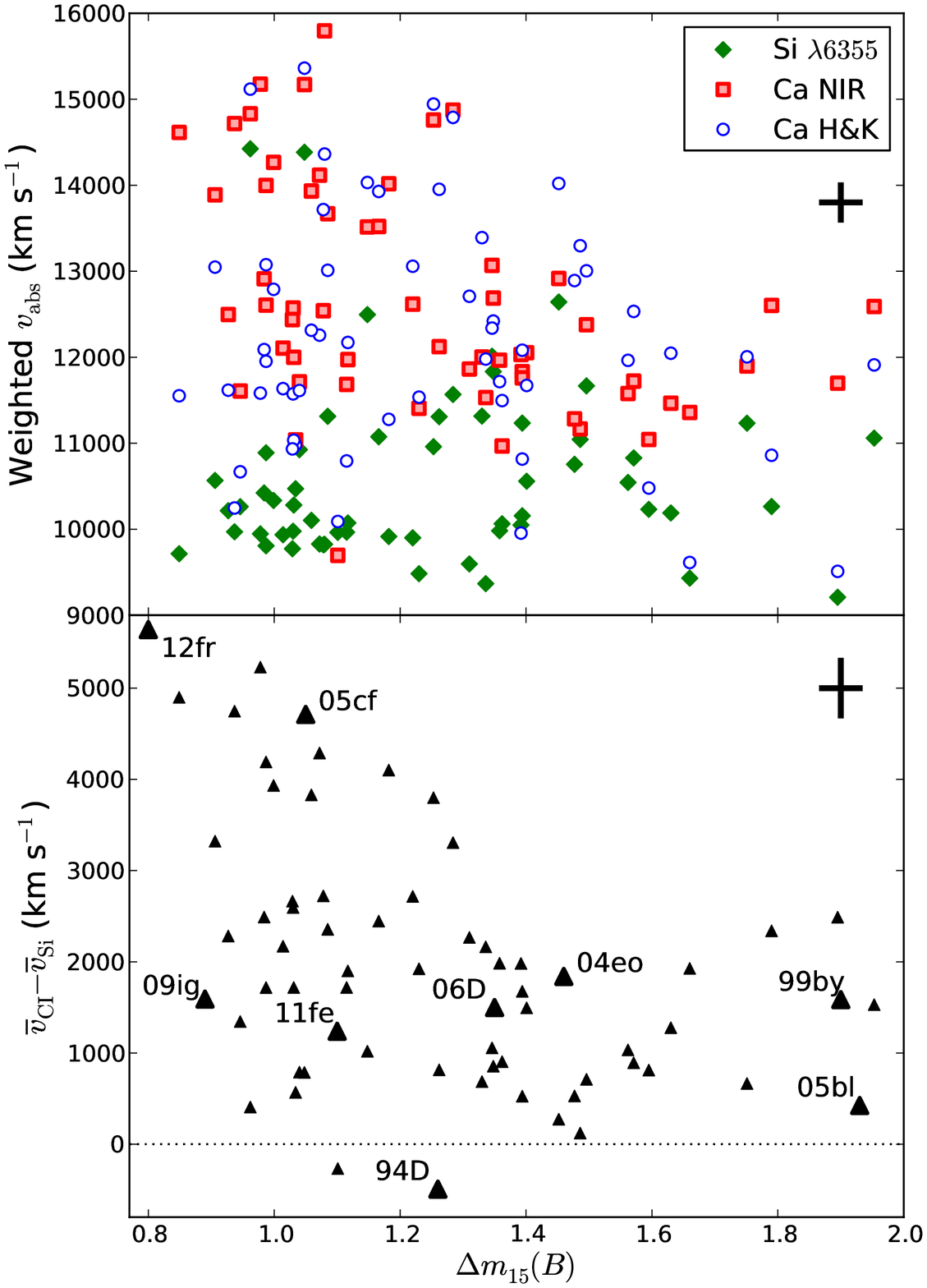}
\caption{Top: Absorption-weighted velocities for \siline, the \canir, and \cahk\ (respectively \vbsi, filled green diamonds; \vbci, shaded red squares; \vbch, open blue circles) versus light-curve decline rate \dmft.  Bottom: Difference between \vbci\ and \vbsi\ versus \dmft. \sneia\ with spectral time series analysed in Section~\ref{sec:hvf_evolution} are labeled by name and have larger symbols.}
\label{fig:vels_vs_dm15}
\end{center}
\end{figure}

To confirm that the trend seen in Figure~\ref{fig:hvf_vs_dm15} is not merely an artifact of our profile-fitting procedure, we turn to the absorption-weighted velocities for the \siline\ line and the \canir. In the top panel of Figure~\ref{fig:vels_vs_dm15} we show the absorption-weighted velocities for these two lines, as well as \cahk, plotted against \dmft; the bottom panel of the same figure shows the velocity difference between the \canir\ and the \siline\ line ($\bar{v}_{\rm CI}-\bar{v}_{\rm Si}$). This figure demonstrates that the behaviour observed for the HVF absorption ratio \rhvf\ is observed with the same qualitative features in the model-independent measurement of mean velocity difference. Rapidly declining ($\Delta m_{15}(B) \geq 1.4$ mag) \sneia\ generally have \canir\ profiles with similar velocities to the \siline\ profile, indicating no HVF influence. Conversely, lower \dmft\ \sneia\ tend to have \canir\ profiles with significantly higher mean velocities than the \siline\ profiles, indicating the presence of HVFs in the \canir.

We note that nearly all the \sneia\ in our sample have higher absorption-weighted velocities in the \canir\ than in the \siline\ line. Even those \sneia\ which have no evidence of HVFs in the \canir\ profile fitting show a slightly higher absorption-weighted velocity by up to $\sim$2,000~\kms. This could be caused either by very weak HVFs or by contamination from other absorption features in this range. We believe a strong candidate for this behaviour is contamination of the \canir\ by the weak \SII\ $\lambda8315$ line, which we identified in the late spectra of SN~2011fe and SN~2004eo (see Section~\ref{sec:hvf_evolution}). Though this line is weak, it presents a wavelength equivalent to the \canir\ at a velocity of 8,800~\kms, meaning even a small amount of absorption by this line will increase the absorption-weighted \canir\ velocity by a large amount. With sufficient S/N, this line is not erroneously fitted as a HVF component of the \canir\ in our profile fitting, but naturally it cannot be removed from the absorption-weighted velocity calculation.

In general, we found the absorption-weighted \cahk\ velocity (\vbch) to be equal to or higher than that of the \canir. In principle, the HVF-to-photospheric pEW ratio for \cahk\ need not be the same value as for the \canir, as the \cahk\ to \canir\ absorption-strength ratio is temperature dependent and the HVF material could be at a different temperature than the photosphere. The \CaII\ lines are in the optically thick regime, meaning these temperature effects are smaller and we should expect the HVF-to-photospheric pEW ratio to be very similar for \cahk\ and the \canir. Thus subtle differences in the absorption-weighted velocities of \cahk\ and the \canir\ are likely due to the influence of \siblue\ on the \cahk\ line complex. Quantifying this effect would likely require a model-based prediction of the \siblue\ absorption strength \citep[perhaps from a $R_{\rm Si}$-based temperature estimate, see discussion in][]{foley12}, which is beyond the intended scope of this work. For these reasons we focus on the \canir, which offers a clean line complex for measuring HVF strength without the influence of other absorption features.

\subsection{HVFs and Light-Curve Colours}
\label{sec:hvf_vs_color}
Next we examine the relationship between HVF strength and the colours of \snia\ light curves. This is particularly important because it has been proposed that the absorption velocity \citep{foley11, fk11} or equivalent width \citep{chotard11} of the \cahk\ line could be driven by the ``intrinsic'' colours (i.e., before reddening by dust) of \sneia. We showed above (Section~\ref{sec:hvf_vs_dmft}) that the velocity profile of \CaII\ varies strongly with decline rate because of the presence of HVFs \citep[consistent with the interpretation of the decline-rate dependence of the \cahk\ profile proposed by][]{maguire12}. This raises the question of whether HVFs could be driving some (or all) of the \snia\ intrinsic colour variability.

\begin{figure}
\begin{center}
\includegraphics[width=0.45\textwidth]{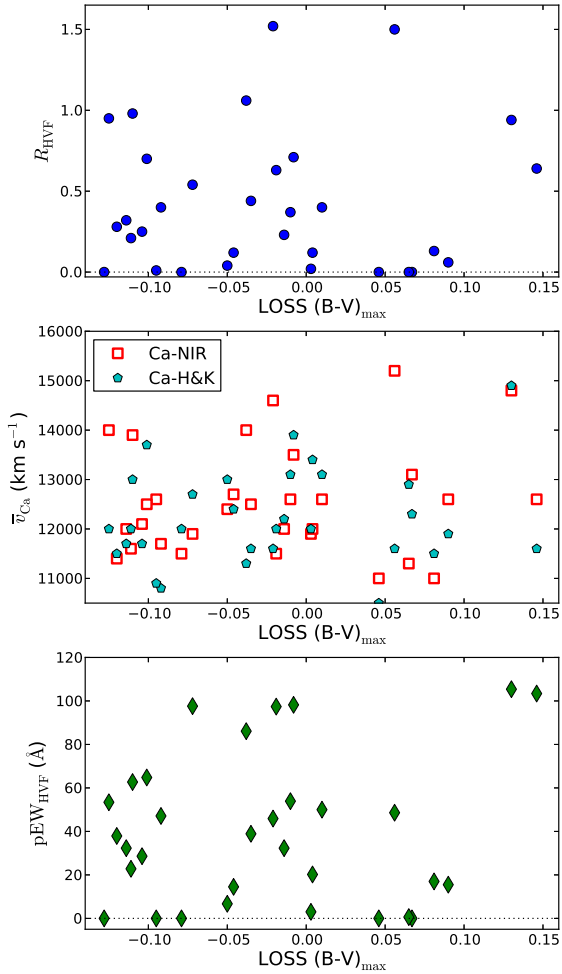}
\caption{Top: HVF strength \rhvf\ versus the $(B-V)$ colour at maximum light. Middle: Absorption-weighted \CaII\ velocities from both the \cahk\ (filled cyan circles) and \canir\ (open red squares) versus colour. Bottom: Absorption equivalent width (pEW) of the HVF component of the \canir\ versus colour.}
\label{fig:hvf_vs_color}
\end{center}
\end{figure}

To test this possibility, we utilise the rest-frame $(B-V)$ maximum-light colours as measured by \citet{ganesh10} from the LOSS \snia\ light curves obtained with the 0.76\,m Katzman Automatic Imaging Telescope \citep[KAIT;][]{filippenko01}. In Figure~\ref{fig:hvf_vs_color} we plot three quantities of interest -- the HVF strength ratio \rhvf, the absorption-weighted mean velocities of \cahk\ and the \canir, and the absorption equivalent width of the HVF component -- against the maximum-light colours of the LOSS sample. Specifically, we focus on those \sneia\ which do {\em not} suffer from substantial reddening by dust in their host galaxy, which we define as $(B-V) \leq 0.15$ mag, in order to gauge whether the blue edge of the distribution has some dependence on an HVF property. This colour cut is slightly more stringent than values typically employed for cosmological analyses \citep[e.g., $c \leq 0.25$;][]{sullivan11a}, but is roughly equivalent to the exponential reddening scale measured by \citet{jha07} and thus is likely to isolate \sneia\ with low reddening.

Examination of the HVF pEW ratio \rhvf\ or the equivalent width of the HVF absorption feature compared to the observed colours of the \sneia\ in our sample shows no obvious correlation of colour with HVF strength.
Similarly, we find no clear correlation of the \CaII\ velocity (either \cahk\ or the \canir) with HVF strength.
These efforts are unfortunately limited by the effect of foreground dust on the observed colours of the \sneia. To fully quantify the effect of HVFs on colours of \sneia, we require a larger sample of \sneia\ which are either confirmed to have no (or low) foreground reddening by dust, or whose dust reddening can be robustly measured from broad wavelength coverage including the NIR. However, our analysis indicates that HVFs do not have a strong effect on the intrinsic colours of \sneia.

\subsection{HVFs and Silicon Velocities}
\label{sec:hvf_vs_vsi}
\begin{figure}
\begin{center}
\includegraphics[width=0.45\textwidth]{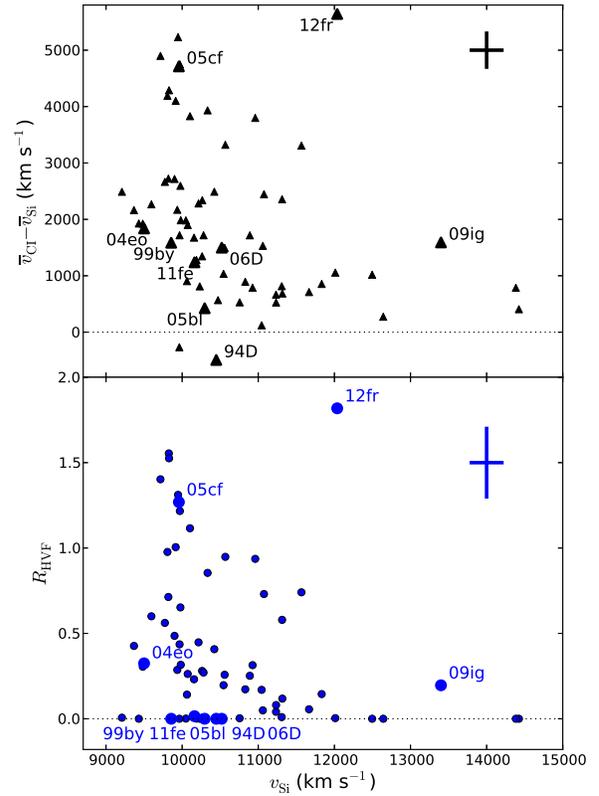}
\caption{Top: Difference in absorption-weighted velocities ($\bar{v}_{\rm CI}-\bar{v}_{\rm Si}$) versus velocity of the \siline\ line at maximum light (\vsi). Bottom: HVF pEW ratio (\rhvf) versus \vsi. In both panels, \sneia\ with spectral time series analysed in Section~\ref{sec:hvf_evolution} are labeled by name and have larger symbols.}
\label{fig:params_vs_vsi}
\end{center}
\end{figure}

Finally we consider the HVF strength compared to the absorption velocity of the \siline\ line at maximum light, \vsi. In Figure~\ref{fig:params_vs_vsi} we show the weighted absorption velocity difference ($\bar{v}_{\rm CI}-\bar{v}_{\rm Si}$) and HVF absorption ratio (\rhvf) plotted against the maximum-light velocity of the \siline\ line. For these plots we use the fitted velocity centre from the \siline\ profile fits of Section~\ref{sec:profile_fits}, but found negligible differences between these values and either the weighted absorption velocity \vbsi\ or the velocity at the absorption minimum.

Here we found a significant dearth of HV \sneia\ with HVFs. More directly, the weighted mean velocity of the \siline\ line and the \canir\ are consistent in the maximum-light spectra of HV \sneia. An example of this behaviour, the HV but slowly declining SN~2002cd ($\Delta m_{15}(B) = 0.96$ mag, $v_{\rm Si} = 14600$ \kms), is shown in the lower panel of Figure~\ref{fig:hvf_profile_fits}. One could posit that this behaviour may indicate that both the \siline\ line and the \canir\ are affected by HVFs, but this notion is not supported by the data for the following reasons. For \sneia\ with \siline\ HVFs it has been observed that the \canir\ HVF is much stronger than the \siline\ HVF [e.g., in SN~2005cf \citep{garavini07, wang05cf}, SN~2009ig \citep{foley09ig}, and SN~2012fr \citep{childress12fr}]. As we will show in Section~\ref{sec:hvf_evolution}, the net result of this behaviour is that the weighted mean velocity difference ($\bar{v}_{\rm CI}-\bar{v}_{\rm Si}$) is still high when HVFs are operative, even for \sneia\ with strong \siline\ HVFs. Thus, a consistent velocity between these two lines would be inconsistent with the observed behaviour of HVFs, but instead indicates a lack of HVFs in either line.

We note that the broad phase range for our sample ($-5 \leq \phi \leq 5$) could result in marginal differences between the \vsi\ we measure and the true \vsi\ at maximum light. However, even for the most rapidly evolving \sneia\ with $\dot v_{\mathrm Si} \approx 100$~\kms~day$^{-1}$ \citep{benetti05}, this would yield a \vsi\ uncertainty of 500\,\kms, which is less than 10\% of the observed range of velocities ($9000 \leq$ \vsi $\leq 15,000$~\kms). Furthermore, the phase coverage of our sample is completely uncorrelated with \dmft, so we expect no bias of \vsi\ with \dmft\ in our sample. Thus, we expect that our phase window has negligible effects on our conclusions.

\subsection{HVFs and the Relationship Between \dmft\ and \vsi}
\label{sec:vsi_vs_dm15_vs_hvf}
The \snia\ expansion velocity, as tracked by \vsi, shows no apparent correlation with the light-curve width \dmft\ (see the top panel of Figure~\ref{fig:vels_vs_dm15}). For our full sample, the Pearson correlation coefficient for \vsi\ and \dmft\ is $r=0.001$, indicating no correlation. However, we found that \sneia\ with high \vsi\ or high \dmft\ (i.e., rapidly declining \sneia) tend to be free of HVFs. We thus examine the relationship between these two quantities when splitting the sample by HVF strength. 

\begin{figure}
\begin{center}
\includegraphics[width=0.45\textwidth]{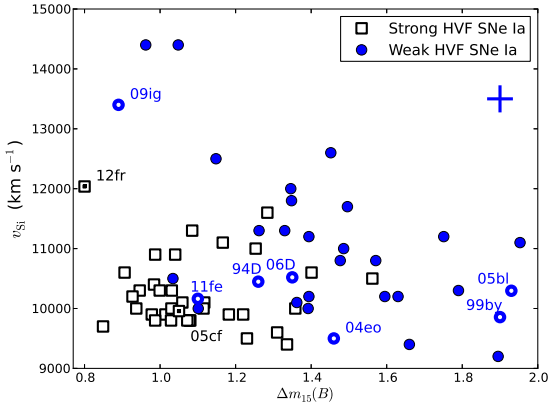}
\caption{\siline\ absorption velocity (from profile fitting) versus light-curve decline rate \dmft\ for weak-HVF \sneia\ (filled blue circles) and strong-HVF \sneia\ (open white squares). The cutoff between weak- and strong-HVF \sneia\ is set here at \rhvf$=0.2$. \sneia\ with spectral time series analysed in Section~\ref{sec:hvf_evolution} are labeled by name and are distinguished by a central dot.}
\label{fig:vsi_vs_dm15_vs_hvf}
\end{center}
\end{figure}

In Figure~\ref{fig:vsi_vs_dm15_vs_hvf} we plot \vsi\ versus \dmft\ for our sample, with points colour coded by HVF strength. We define strong-HVF \sneia\ as those having \rhvf$>0.2$, and analogously those with \rhvf$<0.2$ are classified as weak-HVF \sneia. A striking correlation between \vsi\ and \dmft\ appears evident for the weak-HVF \sneia. The Pearson correlation coefficient for the full weak-HVF sample is $r=0.56$, and if we exclude the extreme outliers (SN~2001fe, SN~2005na, and SN~2011fe), this increases to $r=0.74$. In the three-dimensional space defined by \vsi, \dmft, and \rhvf, \sneia\ appear to congregate in a nearly planar fashion. The \vsi--\dmft\ correlation observed in HVF-weak \sneia\ corresponds to where this plane intersects the \rhvf$=0$ surface.

While the source of this correlation remains unclear, it presents an intriguing result in the context of \snia\ explosions. \sneia\ with high \dmft\ (fast decliners) are generally believed to be less energetic and produce less \nifs, and we see that they have both low photospheric ejecta velocities and no HVFs. Low \dmft\ \sneia\ (slow decliners) are more energetic and produce more \nifs, and we find that they tend to produce {\em either} high photospheric ejecta velocities (i.e., high \vsi) {\em or} strong HVFs, but not both.

The mechanism which sets the observed value of \vsi\ in a given \snia\ is a topic of rigorous study. Many detonation models -- either sub-Chandra double detonations \citep{sim10} or delayed detonations \cite[e.g.,][]{blondin13,seitenzahl13} --
generically result in stratified ejecta where \nifs\ is produced in deeper layers than IMEs like silicon. The consequence of this stratification is that \vsi\ effectively traces the boundary between IMEs and \nifs, so that a more energetic \snia\ which produces more \nifs\ naturally has a higher \vsi\ because the IME-\nifs\ boundary is at larger radius. The fact that high explosion energy (i.e., low \dmft) is not always accompanied by high \vsi\ could be interpreted as an observational challenge to this picture. However, we have shown here that in cases without HVFs the observations agree qualitatively with this stratification prediction for the behaviour of \vsi. Why HVF-strong \sneia\ 
deviate from this trend remains to be seen, but we point out that the strongest-HVF \snia, SN~2012fr, was shown to be highly stratified \citep{childress12fr}, so a lack of stratification seems unlikely to provide a simple explanation for this behaviour.

\section{HVF Evolution for Extensively Observed \sneia}
\label{sec:hvf_evolution}
Our investigation of maximum-light spectra showed an apparent dearth of HVFs in the maximum-light spectra of \sneia\ with either rapidly declining light curves or high photospheric velocities. However, it has been previously observed that HVFs are stronger at earlier epochs in \sneia\ \citep{mazzali05b}, so we will examine the evolution of HVFs in \sneia\ having extensive spectroscopic coverage to inspect whether HVFs are ubiquitous in the earliest spectra of \sneia. To perform this analysis, we downloaded spectral time series for several \sneia\ spanning a range of \dmft, mostly via the WISEREP \citep{wiserep} repository. In Table~\ref{tab:time_series_sne} we summarise the \snia\ sample, light-curve decline rate, and literature sources for the spectra. We note that this sample is not meant to be a comprehensive collection of all \snia\ spectral data, but instead was chosen to have broad sampling in \dmft\ with \sneia\ having extensive spectral time series.

\begin{table*}
\begin{center}
\caption{Time Series \sneia\ and their Maximum-Light Properties}
\label{tab:time_series_sne}
\begin{tabular}{llrrrll}
\hline
SN        & \dmft\ & \vsi\ & \rhvf\ & \vbci\ $-$ \vbsi\ & LC Ref. & Spec. Ref. \\
          & (mag)  & (km\,s$^{-1}$) & & (km\,s$^{-1}$) & & \\
\hline
SN~2012fr & 0.80   & 12000 & 1.82 & 5600 & 1  & 2 \\
SN~2009ig & 0.89   & 13400 & 0.19 & 1600 & 3  & 3 \\
SN~2005cf & 1.05   & 10000 & 1.27 & 4700 & 4  & 4,5 \\
SN~2011fe & 1.10   & 10200 & 0.02 & 1200 & 6  & 6,7 \\
SN~1994D  & 1.26   & 10400 & 0.00 & -500 & 8  & 8, 9, 10 \\
SN~2006D  & 1.35   & 10500 & 0.00 & 1500 & 11 & 12 \\
SN~2004eo & 1.46   &  9500 & 0.32 & 1800 & 13 & 13 \\
SN~1999by & 1.90   &  9900 & 0.00 & 1600 & 14 & 14 \\
SN~2005bl & 1.93   & 10300 & 0.00 &  400 & 15 & 15 \\
\hline
\end{tabular}
\end{center}
{\bf References:} 
(1) Contreras et al., 2013, in preparation;
(2) \citet{childress12fr};
(3) \citet{foley09ig};
(4) \citet{wang05cf};
(5) \citet{garavini07};
(6) \citet{pereira13};
(7) \citet{parrent12};
(8) \citet{patat96};
(9) CfA \citep{blondin12};
(10) BSNIP \citep{bsnip1};
(11) \citet{hicken09a};
(12) \citet{rcthomas07};
(13) \citet{pastorello07};
(14) \citet{garnavich04};
(15) \citet{taubenberger08}.
\end{table*}

For the extensively observed \sneia\ analysed here, we repeat our profile-fitting procedure of Section~\ref{sec:profile_fits}, again taking advantage of the \siline\ line to measure the photospheric velocity profile and constrain the fit of the HVF in the \canir. However, some \sneia\ in this set exhibit HVFs in the \siline\ line at very early epochs (though it has consistently faded by maximum light), making it more challenging to definitively identify the photospheric velocity profile. To mitigate this, we fit the \siline\ line as a double-Gaussian profile in some of the earliest epochs of SN~2012fr \citep[already presented by][]{childress12fr}, SN~2005cf, and SN~2009ig.

\begin{figure}
\begin{center}
\includegraphics[width=0.45\textwidth]{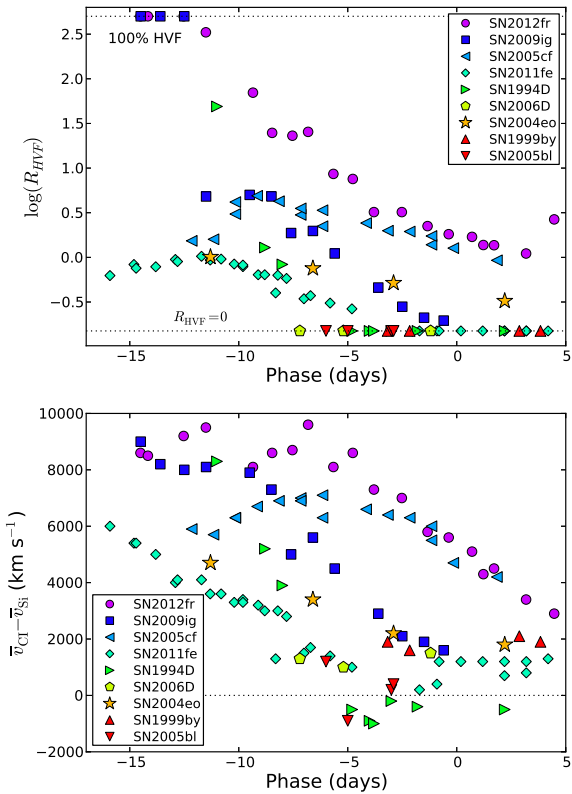}
\caption{Top: HVF pEW ratio (\rhvf) versus phase for a sample of well-observed \sneia. Bottom: Difference in absorption-weighted velocities ($\bar{v}_{\rm CI}-\bar{v}_{\rm Si}$) versus phase for the same sample.}
\label{fig:time_series_hvf}
\end{center}
\end{figure}

For all epochs of all \sneia, we calculate the HVF strength \rhvf\ in the \canir, as well as the absorption-weighted velocity of \siline\ (\vbsi) and the \canir\ (\vbci). In Figure~\ref{fig:time_series_hvf} we plot these quantities against phase for the SNe in our sample, with colour coding ordered by \dmft.

The first apparent conclusion from inspection of Figure~\ref{fig:time_series_hvf} is that HVF evolution does not appear to follow any universal pattern in \sneia.  Similarly, the HVF strength at maximum light does not provide an unambiguous prediction for the pre-maximum-light behaviour of the HVFs. We briefly describe the behaviour of HVFs for each SN in our sample.

{\it SN~2012fr}. SN~2012fr is the most slowly declining (\dmft\ $=0.80$ mag) \snia\ in our sample, and also has the most extreme HVFs. As noted by \citet{childress12fr}, the \canir\ is completely dominated by HVFs during the first few days, and then the HVFs decline in strength but remain very strong overall (compared to the other \sneia) even through maximum light.

{\it SN~2009ig}. This slowly declining (\dmft\ $=0.89$ mag) \snia\ also begins with very strong HVFs that dominate the \canir. However, the HVFs decline in strength much more rapidly than in SN~2012fr and have faded to nearly zero by about maximum light. This behaviour marks an interesting contrast to SN~2012fr (which has a similar luminosity and decline rate) because SN~2009ig is a firm member of the HV class \citep[\vsi = 13,400 \kms\ at maximum light; ][]{marion13}, which we noted above tend to have no HVFs at maximum light.

{\it SN~2005cf}. Dubbed the ``golden standard'' \snia\ \citep{wang05cf}, this \snia\ has a very typical decline rate and silicon velocity at maximum light. It exhibits very strong HVFs, and has perhaps the slowest HVF-strength decline rate of all \sneia\ considered here. The origin of the persistently strong HVFs (compared, especially, to SN~2011fe) is an intriguing question.

{\it SN~2011fe}. The closest and best-observed \snia\ in a generation, SN~2011fe was a prototypical example of a normal \snia. Its earliest epochs show no HVF in the \siline\ line, but clear HVFs of modest strength in the \canir. The HVF strength shows a small but statistically significant increase for the first week after the SN explosion, and then a gradual decline to zero a few days before maximum light.

{\it SN~1994D}. Considered the prototypical ``core-normal'' \snia, SN~1994D shows strong HVFs in its first few spectroscopic epochs, which decline in strength at a rate more rapid than that of any other SN in our sample. SN~1994D has a normal \vsi\ at maximum light \citep[$\sim$ 10,700 \kms;][]{bsnip2}, so this rapid HVF decline is apparently not associated with HV behaviour.

{\it SN~2006D}. This \snia\ showed some of the strongest absorption by unburned carbon ever observed in a \snia\ \citep{rcthomas07}, and has a modestly rapid decline rate (\dmft\ $=1.35$ mag). Its earliest epoch of observations is $-7$ days, and it shows no evidence for HVFs in any of its spectra. Whether the lack of HVFs is related to the presence of carbon or is simply due to the rapid decline rate is unclear. It is worth noting that one of the strongest HVF \sneia, SN~2005cf, was also highlighted by \citet{rcthomas11} for having a very clear carbon absorption feature in its early spectra (before about $-7$ days).  Carbon is presumed to originate from the surface of the exploding WD, and it is possible that HVFs could also arise from this same physical location. Thus, we believe that the relationship between unburned carbon and HVFs is a topic worthy of further exploration. Since both spectroscopic behaviours are strongest at earliest epochs, this underscores the need for timely discovery and classification of young \sneia.

{\it SN~2004eo}. This rapidly declining (\dmft\ $=1.46$ mag) \snia\ may be at the very faint end of the distribution of normal \sneia\ \citep{pastorello07}. Interestingly, SN~2004eo shows a fairly strong \canir\ HVF ($R_{\rm HVF} \approx 1$) in its first epoch ($-11$ days), which declines in strength very slowly. We note here that in the final two epochs ($-3$ and $+2$ days), the fit to the HVF may be influenced by the weak \SII\ $\lambda8315$ line. This feature was also evident in SN~2011fe spectra starting at around $-4$ days, but due to the high S/N of those spectra fitting of that feature as an HVF of \canir\ resulted in a poor \chisq, meaning that for SN~2011fe the profile fits were not affected by this line. For the medium-S/N spectra of SN~2004eo this \SII\ line is more problematic, and we can see that the profile was fitted as an unnaturally strong HVF with the photospheric \canir\ having a lower velocity than the \siline\ line (which was generally not the case for the \canir\ profiles that were well fit). Thus, for the last two epochs here we suspect the true \rhvf\ value to be closer to 0. This underscores the importance of also calculating the model-independent absorption-weighted velocities: in this quantity we find a leveling off near maximum light, similar to SN~2011fe, consistent with the hypothesis that HVFs in SN~2004eo have actually faded by maximum light. At the very least, the clear presence of HVFs in the earliest epochs ($-11$ and $-7$ days) may point to the ubiquity of HVFs in all normal \sneia.

{\it SN~1999by and SN~2005bl}. These extremely rapidly declining \sneia\ (\dmft\ $=1.90$ and 1.93 mag for SN~1999by and SN~2005bl, respectively) are members of the subclass of peculiar \sneia\ spectroscopically similar to the subluminous SN~1991bg \citep{filippenko91bg,leibundgut91bg}. These SNe have the earliest spectra ($-4$ and $-6$ days) of any members of the subclass, but show no evidence for HVFs in any of their spectra. This is unsurprising given the absence of HVFs by this epoch in more-slowly declining \sneia\ and the apparent trend of HVF strength with \dmft. Whether SN~1991bg-like \sneia\ have HVFs at earlier epochs remains to be seen, and requires earlier detections of these very subluminous SNe.

Using a sample of \sneia\ spanning decline rates from \dmft$=0.97$ mag \citep[SN~2003kf;][]{hicken09a} to \dmft$=1.28$ mag \citep[SN~2002er;][]{ganesh10}, \citet{mazzali05b} found that HVFs were evident in their entire sample of \snia\ spectra taken about a week before maximum light. We confirm this behaviour here for an extended range of \dmft\ (0.80 mag for SN~2012fr to 1.46 mag for SN~2004eo), and echo the suggestion by \citet{mazzali05b} that HVFs appear to be a ubiquitous feature of normal \sneia\ which are uniformly stronger at earlier epochs.

However, we find that the pre-maximum HVF evolution does not follow a consistent pattern among all \sneia. A consequence of this is that the relative ranking of HVF strength in a sample of \sneia\ at maximum light is not always preserved at very early epochs. Thus, a larger set of pre-maximum \snia\ spectra appears necessary to fully categorise this behaviour and define the best metric(s) to describe the HVF strength of a given \snia. Another key result is that several \sneia\ showing no HVFs at maximum light (specifically the high-\vsi\ SN~2009ig and the moderately fast-declining SN~2004eo) {\em do} exhibit HVFs at earlier epochs. Again, this suggests that the HVF strength at maximum light is not a comprehensive metric of HVF strengths in \sneia, as those which are HVF-free at maximum light may have some diversity at earlier epochs when the HVF strength is nonzero.

\section{Discussion and Conclusions}
\label{sec:conclusions}
In this work we conducted an investigation of HVFs in the maximum-light spectra of a sample of 58 \sneia, as well as HVFs at all pre-maximum-light epochs in several \sneia\ having extensive spectral time series. Specifically, we focused on HVFs in the \canir\ feature, using epochs where the \siline\ line was clean of HVFs to help define the photospheric velocity profile. We quantified the HVF strength with two metrics: (1) the pEW ratio of the \canir\ HVF to the photospheric absorption component, as measured from a fit to the \canir\ velocity profile; and (2) the difference in the absorption-weighted velocity of the \siline\ line and the \canir. The first method provided a physically meaningful but model-dependent measurement of the HVF strength, while the second allowed for agnostic measurement of the mean velocity of the \siline\ line and \canir\ absorption profiles. From our analysis, we found important trends in HVF behaviour, as follows. 
\begin{itemize}
  \item \sneia\ with rapidly declining light curves ($\Delta m_{15}(B) \geq 1.4$ mag) have consistently no HVFs at maximum light, and for normal \sneia\ the average HVF strength tends to increase with decreasing light-curve decline rate.
  \item \sneia\ with a high silicon velocity (\vsi $\geq$~12,000~\kms) at maximum light \citep[i.e., the ``HV'' subclass identified by][]{wang09} also show no evidence for HVFs at maximum light. A trend of HVF strength with \vsi\ is also evident, but it is weaker than the trend with \dmft.
  \item The pre-maximum-light evolution of HVFs exhibits significant diversity, with the rate of HVF fading showing no clear dependence on the overall HVF strength, \dmft, or \vsi.
  \item Some \sneia\ which show weak or no HVFs at maximum light (specifically, the high-\vsi\ SN~2009ig and the fast-declining SN~2004eo) were found to have HVFs in very early spectra.
\end{itemize}

The original goal of this project was to examine the influence of HVF Ca on the \cahk\ line complex, to confirm whether the stretch dependence of this line complex noted by \citet{maguire12} was caused by HVF Ca as those authors proposed, or whether it could be due primarily to the \siblue\ line as posited by \citet{foley12}. We found that although the wavelength-space degeneracy of HVF \cahk\ and \siblue\ prevented a direct decoupling of these lines in the \cahk\ line complex, the \canir\ provided a clean independent measurement of the influence of HVFs. Our finding that HVFs in the \canir\ are stronger in more-slowly declining (i.e., higher ``stretch'') \sneia\ supports the interpretation of \citet{maguire12}. However, the \siblue\ line does appear to impact the \cahk\ line complex in nearly all cases, as evidenced by the difference in absorption-weighted velocities for \cahk\ compared to the \canir, lending some credence to the argument of \citet{foley12} that \siblue\ affects the \cahk\ line profile.

The origin of the HVFs remains unknown, but numerous hypotheses have been presented (see Section~\ref{sec:intro}). Our findings provide important observations of the behaviour of HVFs in relation to \dmft\ and \vsi\ which offer constraints on possible models of their origin. While the HVFs themselves are identified by their distinct absorption profiles, the HVF material could potentially cause some smooth alteration of the full SN spectral energy distribution not easily identifiable from SN spectra, which in turn could impact the degree to which \snia\ luminosities can be standardized. We confirmed the findings of previous authors that the strength of HVFs is consistently greater at earlier epochs. This underscores the need for the earliest possible spectroscopic observations of \sneia\ in order to further inspect this important facet of their behaviour.



\vskip11pt
{\em Acknowledgements:}
We thank Stefan Taubenberger for providing his spectra of SN~2005bl; Jeffrey Silverman for his hard work on the BSNIP sample; the late Weidong Li for his key role in obtaining the KAIT light curves published by LOSS \citep[][]{ganesh10}; Stuart Sim, Richard Scalzo, Brad Tucker, and Ryan Foley for helpful discussions; and the Lick Observatory staff for their assistance with the observations. We also thank the anonymous referee for thoughtful and constructive comments.
This research was conducted by the Australian Research Council Centre of Excellence for All-sky Astrophysics (CAASTRO), through project number CE110001020. B.P.S. acknowledges support from the Australian Research Council Laureate Fellowship Grant LF0992131. 
A.V.F. is grateful for the generous financial support of NSF grant AST-1211916, the TABASGO Foundation, and the Christopher R. Redlich Fund. KAIT has been funded by donations from Sun Microsystems, Inc., the Hewlett-Packard Company, AutoScope Corporation, Lick Observatory, the NSF, the University of California, the Sylvia \& Jim Katzman Foundation, the Christopher R. Redlich Fund, the Richard and Rhoda Goldman Fund, and the TABASGO Foundation.
This research has made use of NASA's Astrophysics Data System (ADS).

\bibliographystyle{apj}
\bibliography{hvf_sneia}

\end{document}